\documentstyle{article}

\def\1ad{\mbox{\normalsize $^1$}}
\def\2ad{\mbox{\normalsize $^2$}}
\def\3ad{\mbox{\normalsize $^3$}}
\def\4ad{\mbox{\normalsize $^4$}}
\def\5ad{\mbox{\normalsize $^5$}}
\def\6ad{\mbox{\normalsize $^6$}}
\def\7ad{\mbox{\normalsize $^7$}}
\def\8ad{\mbox{\normalsize $^8$}}
\def\makefront{\vspace*{1cm}\begin{center}
\def\newtitleline{\\ \vskip 5pt}
{\Large\bf\titleline}\\
\vskip 1truecm
{\large\bf\authors}\\
\vskip 5truemm
\addresses
\end{center}
\vskip 1truecm
{\bf Abstract:}
\abstracttext
\vskip 1truecm}
\def\inbar{\vrule height1.5ex width.4pt depth0pt}
\def\IC{\relax\,\hbox{$\inbar\kern-.3em{\rm C}$}}
\def\dop{{\rm d}\hskip -1pt}

\def\IR{\relax{\rm I\kern-.18em R}}

\begin {document}
{\hfill{KUL-TF-98/5, DFTT-2/98,
 IFUM/605-FT, hep-th/9801140}}
\def\titleline{Special geometry of Calabi-Yau compactifications 
\newtitleline near a rigid limit\footnote{To appear 
in the proceedings of {\it Quantum aspects of 
gauge theories, supersymmetry and unification}, 
Neuchatel University, 18-23 September 1997. Talk presented by F.D.}}

\def\authors{Marco Bill\'o \1ad,
Frederik Denef \1ad \4ad,
Pietro Fr\`e \2ad, \\
Igor Pesando \2ad,  Walter Troost \1ad \5ad,\\
Antoine Van Proeyen \1ad \6ad
and  Daniela Zanon \3ad }

\def\addresses{  \1ad
Instituut voor theoretische fysica, \\
Katholieke Universiteit Leuven, B-3001 Leuven, Belgium\\
\2ad Dipartimento di Fisica Teorica dell' Universit\`a,\\
                 via P. Giuria 1,
                I-10125 Torino, Italy\\
\3ad Dipartimento di Fisica dell'Universit\`a di Milano and\\
INFN, Sezione di Milano, via Celoria 16,
I-20133 Milano, Italy \\
\4ad Aspirant FWO,
\5ad Onderzoeksleider FWO,
\6ad Onderzoeksdirecteur FWO.}

\def\abstracttext{We discuss, in the framework of special K\"{a}hler geometry,
some aspects of the ``rigid limit'' of type IIB string theory
compactified on a Calabi-Yau threefold. We outline the general idea
and demonstrate  by direct analysis of a specific example how this limit
is obtained. The decoupling of gravity and the reduction of
special K\"{a}hler geometry from local to rigid is demonstrated explicitly,
without first going to a noncompact approximation of the Calabi-Yau.
In doing so, we obtain the Seiberg-Witten Riemann surfaces corresponding to
different rigid limits as degenerating branches
of a higher genus Riemann surface,  defined for all values of the moduli.
Apart from giving a nice
geometrical picture, this allows one to calculate easily some gravitational corrections
to e.g. the Seiberg-Witten central charge formula.
We make some connections to the 2/5 brane
picture, also away from the rigid limit, though only at the formal level.}

\makefront

\section{Introduction}
String theory has proven to be a quite valuable tool in obtaining exact
results for nontrivial supersymmetric quantum field theories.
In many of these solutions, special K\"{a}hler geometry,
and in particular the extraction
of its rigid limit, plays a key role. Indeed, special geometry is rich enough
to control
the Coulomb branch of moduli space for $N=2$, $D=4$ theories,
and hence their two derivative effective action, while in
many cases it is also sufficiently constrained to allow for exact solutions.

There are two types of special K\"{a}hler geometry: ``rigid'' and ``local''. Local
special geometry applies to local supersymmetry, i.e. supergravity and
string theory, and rigid special geometry to rigid supersymmetry, hence to
supersymmetric gauge theories in flat spacetime.
Thus to extract pure quantum field theory results from string theory, one has to
``turn off'' gravity and  go to a rigid limit of local special geometry.

Type IIB string theory compactified on a Calabi-Yau manifold gives an $N=2$
theory in four dimensions. The (local) special geometry of the vector multiplet
moduli space in this
case is given by the classical geometry of the CY complex structure moduli,
and is known to receive no quantum corrections. Therefore, by going to
a rigid limit of this classical moduli space and identifying the corresponding
rigid low energy quantum theory (usually a field theory),
one should get an exact solution for the two derivative low energy
effective action of this theory \cite{KKLMV,KLMVW}.
The identification of the rigid theory is conceptually the most
nontrivial part, and indeed this was only solved after the discovery
of heterotic/type II duality (suggesting nonabelian gauge theories)
and D-branes as solitonic states (providing the ''missing'' massive
gauge vector multiplets).

In many cases, this procedure reduces the local special geometry of the
Calabi-Yau moduli space to the rigid special geometry of the moduli space of a
certain class of Riemann surfaces,
reproducing and extending the Seiberg-Witten solution of $N=2$
quantum Yang-Mills theory. Furthermore, many features of quantum field theory
have a beautiful geometrical interpretation in this framework, and this provides
quite elegant solutions to problems which would be hard to tackle with
 ordinary field theory techniques, like for example the existence
and stability of BPS states \cite{KLMVW,Klemmreview,Lerchereview}.

By now, a very large class of $N=2, d=4$ quantum field theories
(and even more exotic theories) can be  ``engineered'' and solved geometrically in this
way. The usual procedure \cite{GE,GE-review}
is to find a local IIA model which in the rigid limit produces the field theory
to be solved; to map this IIA theory to an equivalent IIB theory using
local mirror symmetry; and finally to solve this IIB theory exactly (in the low energy
field theory limit) using classical geometry. One argues that the restriction to local
models and local mirror symmetry --- where ``local'' means that one only considers
a certain small region of the Calabi-Yau manifold --- is allowed roughly
because the relevant
(light brane) degrees of freedom are all localized well inside that region.

An alternative, but not unrelated, approach is to make use of M-theory brane
configurations in flat space \cite{M-sol}.

In this paper we study in detail the rigid limit on the type IIB side, without
assuming a priori that we can restrict ourselves to the local considerations
mentioned above. We first
discuss some general aspects of special geometry and its rigid limit. Then we
show by direct analysis how the $SU(3)$ rigid limit for IIB on the CY manifold
$X^*_{24}[1,1,2,8,12]$ is obtained from the full local special geometry.
The decoupling of gravity and the reduction of
special K\"ahler geometry from local to rigid is demonstrated explicitly.
In doing so, we obtain the Seiberg-Witten Riemann surfaces corresponding to the
different rigid limits as degenerating branches
of a higher genus Riemann surface,
defined for {\em all} values of the Calabi-Yau moduli. Apart from giving a nice
geometrical picture, this allows us to calculate easily some gravitational corrections
to e.g. the Seiberg-Witten BPS mass formula.

Along the way, we make some connections to the brane
picture, though only at the formal level, where it can also be extended
to the global Calabi-Yau description.

More details can be found in \cite{CYK3}.

\section{From local to rigid special geometry}

A local special K\"{a}hler manifold has a K\"{a}hler potential of the form
\begin{equation}
{\cal K} = - \ln (-i v^t q^{-1} \bar{v}),
\label{calK}
\end{equation}
where $v$ is a certain holomorphic section of a symplectic vector bundle
and $q$ an invertible, antisymmetric and constant matrix, the symplectic form.
The symplectic section has to satisfy the following integrability condition:
\begin{equation}
({\cal D} v)^t q^{-1} v = 0, \label{intcond}
\end{equation}
where ${\cal D} = \partial + \partial {\cal K}$.

For type IIB theory compactified on a Calabi-Yau manifold, the vector multiplet
moduli space coincides with the complex structure moduli space.
This space has local special K\"{a}hler geometry with
\begin{eqnarray}
v_\Lambda &=& \int_{C_\Lambda} \Omega \nonumber\\
q_{\Lambda \Sigma} &=& C_\Lambda \cdot C_\Sigma,
\end{eqnarray}
where $\Omega$ is the holomorphic 3-form on the CY, $\left\{ C_\Lambda \right\}_\Lambda$
is a basis of 3-cycles, and the dot denotes the intersection product.
The number of 3-cycles is equal to two plus twice the number of massless
vector multiplets.

A rigid special K\"{a}hler manifold on the other hand has a K\"{a}hler
potential of the form
\begin{equation}
K = i V^t Q^{-1} \bar{V}.
\label{K}
\end{equation}
Again $V$ is a holomorphic section of a symplectic vector bundle,
with symplectic form $Q$, but now the
integrability condition is:
\begin{equation}
(\partial V)^t Q^{-1} V = 0.
\end{equation}

Rigid special geometry is realised on the moduli space of Seiberg-Witten
Riemann surfaces by
\begin{eqnarray}
V_A &=& \int_{c_A} \lambda \nonumber\\
Q_{AB} &=& c_A \cdot c_B,
\end{eqnarray}
where $\lambda$ is the Seiberg-Witten meromorphic 1-form on the Riemann surface,
and $\left\{ c_A \right\}_A$
is a basis of 1-cycles,
as many as twice the number of massless vector multiplets.

Quite generally, a rigid limit of local special geometry can be obtained as follows.
Suppose there is a region in moduli space where we can choose a
subset of coordinates $(u_1,\ldots, u_r)$ and symplectic vector
components $(v_1,\ldots, v_{2r})\equiv V $ such that the
K\"{a}hler potential (\ref{calK}) can be written as
\begin{equation}
{\cal K} = - \ln \left( M^2 - i V^t {Q}^{-1} \bar{V} + R \right),
\label{calK2}
\end{equation}
where $M$ is independent of the $u_i$, ${Q}$ is real, invertible
and antisymmetric, and $R$ is a remainder such that
\begin{equation}
\frac{V^t {Q}^{-1} \bar{V}}{M^2} \to 0,
\;\; \frac{R}{V^t {Q}^{-1} \bar{V}} \to 0
\label{limdef}
\end{equation}
when approaching a certain locus in this region. Then close to this locus,
we can make the following expansions:
\begin{eqnarray}
V &=& V_0 + \ldots \\
{\cal K} &=& - \ln M^2 + \frac{1}{M^2} i V_0^t {Q}^{-1} \bar{V}_0 + \ldots
\label{Kahlerexpansion} \\
{\cal D}_u V &=& \partial_u  V_0 + \ldots
\end{eqnarray}
where the dots indicate subleading terms that can be neglected in the limit
under consideration. Note that (\ref{Kahlerexpansion}) is, up to an irrelevant
$u$-independent term,
precisely the expression for the K\"{a}hler potential
in {\em rigid} special geometry.
Moreover, the integrability condition (\ref{intcond}) reduces to
\begin{equation}
(\partial_u V_0)  {Q}^{-1} V_0 = 0,
\end{equation}
which is precisely the integrability condition defining rigid special geometry.
Thus we find that the geometry of the moduli subspace
parametrized by the moduli $u_i$ and endowed with the symplectic
vector $V_0$, is essentially rigid special K\"ahler.

The limit described above will be called a {\em rigid limit},
and can be thought of as sending the Planck mass to infinity, effectively decoupling
gravity from the degrees of freedom associated with the rigid moduli $u_i$.

\section{An explicit example}

Let us now explicitly demonstrate how the above general structures do
indeed arise in a specific example, namely type IIB string theory
compactified on the Calabi-Yau manifold $X^*_{24}[1,1,2,8,12]$.
Choosing appropriate affine coordinates $(\zeta,x,\xi,y)$ on (a patch of) the
ambient weighted projective space \cite{CYK3}, this CY manifold,
endowed with a generic complex structure, can be represented by
\begin{equation}
y^2 + \frac{1}{2} (\xi + \frac{B^\prime(\zeta)}{\xi}) + P(x) = 0
\label{W_elliptic}
\end{equation}
where
\begin{eqnarray}
B^\prime(\zeta) &=& \frac{B}{2} (\zeta + \frac{1}{\zeta}) + 1 \label{B} \nonumber\\
P(x) &=& x^3 + A_1 \, x + A_2.
\label{P}
\end{eqnarray}
For our purposes we do not need to worry about the points at infinity,
so this equation gives us all the necessary informations about the Calabi-Yau.
The complex structure moduli space has complex dimension three and is parametrized
by the coefficients $(B,A_1,A_2)$.
The holomorphic $3$-form is given by
\begin{equation}
\Omega = \frac{\dop\zeta}{\zeta} \wedge \dop x \wedge \Omega_T
\end{equation}
with
\begin{equation}
\Omega_T = \frac{1}{2 \pi i} \, \frac{1}{y} \frac{\dop\xi}{\xi}\ .
\end{equation}
The Calabi-Yau admits two elliptic fibrations. One
of them is manifest in (\ref{W_elliptic}):
at fixed $(\zeta,x)$, this equation indeed describes a torus.
The holomorphic $1$-form on the elliptic fibre is equal to $\Omega_T$.
The other fibration is obtained by taking $(\zeta,\xi)$ as base space variables.
The first fibration is of practical use in obtaining periods and monodromies
close to a rigid limit, while the second one is more suitable for the large complex
structure limit. Since we are interested in the former, we will only consider
the first fibration.

The torus fibre degenerates when one of its 1-cycles vanishes. This occurs at
a complex codimension~1 locus in the base of the elliptic fibration, that is,
on a Riemann surface $\Sigma$ (or on a 5+1 dimensional submanifold
$M_4 \times \Sigma$ if we consider four dimensional space-time $M_4$ to
be part of the base). $\Sigma$ consists of two components given by the
equations
\begin{eqnarray}
\Sigma_\alpha:&& P(x)^2 - B^\prime(\zeta) = 0 \nonumber\\
\Sigma_\beta:&& B^\prime(\zeta) = 0.
\end{eqnarray}
$\Sigma_\alpha$ is a six sheeted cover of the $\zeta$-plane and has genus 5, and $\Sigma_\beta$ is
a disconnected trivial two-sheeted cover of the $x$-plane (in the CY patch under consideration).
Denote the vanishing torus cycle on $\Sigma_\alpha$ ($\Sigma_\beta$) 
by $\alpha$ ($\beta$). It can be shown \cite{CYK3} that $\{ \alpha, \beta \}$
is a canonical basis of 1-cycles on the torus fibre, that is,
$\alpha \cdot \beta = 1$.

Let us now consider type IIB string theory compactified on this CY space.
In order to calculate the special K\"{a}hler potential on the vector moduli space,
one has to find a basis of
CY 3-cycles together with their intersections and periods.  Furthermore, 
minimal volume
3-branes wrapped around 3-cycles produce BPS states, so an explicit
description of the
relevant 3-cycles will allow us to identify those states.
A suitable description can be obtained by constructing the 3-cycles as
circle fibrations over two dimensional surfaces
in the base manifold, where the circles are nontrivial cycles on the
torus fibre. To obtain a smooth closed 3-cycle, a base surface with fibre
$\alpha$ ($\beta$)
should either be closed or end on $\Sigma_\alpha$ ($\Sigma_\beta$),
and it should not cross $\Sigma_\beta$ ($\Sigma_\alpha$).
It is easy to see that a cylinder stretched between two sheets yields
a 3-cycle with topology $S^1 \times S^2$, while a disc gives an $S^3$ and a
(non-stretched) torus an $S^1 \times S^1 \times S^1$.

The period of a 3-cycle $C$ with fibre $\alpha$ and base surface $S$ can be expressed as
\begin{equation}
\int_C \Omega = \int_S f_\alpha(\zeta,x) \frac{\dop\zeta}{\zeta} \wedge \dop x
\end{equation}
where $f_\alpha$ can be expressed in terms of the complete elliptic integral
${\bf K}(z) = \frac{\pi}{2} F(\frac{1}{2},\frac{1}{2},1;z)$. E.g. for
points sufficiently close to $\Sigma_\alpha$, we have:
\begin{equation}
f_\alpha (\zeta,x)=\int_\alpha\Omega_T= \frac{2\sqrt{2}}{\pi}
 \left(B^\prime(\zeta)\right)^{-1/4} {\bf K}
( \frac{1}{2} (1 + P(x)/\sqrt{B^\prime(\zeta)}) ). \label{fa}
\end{equation}
Analogous expressions hold for cycles with $\beta$ fibre.

One can see the base surfaces $S$ as
2-branes ending on the 5-branes $M_4 \times \Sigma$, thus making contact,
at least formally, with the brane configuration picture \cite{M-sol}.
This (partially) extends the connection made in \cite{KLMVW} in the ALE
approximation, to the general Calabi-Yau case. 
The analogy can be made more precise
by specifying an embedding of the 2- and 5-branes in a suitable 10 or 11
dimensional manifold, but we will not dwell on this here.

Now we turn to the rigid limit. From the physical interpretation of the 3-branes,
we expect to approach such a limit when we tune the moduli such that the
genus 5 Riemann surface $\Sigma_\alpha$ develops a branch of (almost)
coinciding sheets.
Indeed, then we get a set of small volume 3-cycles, namely those constructed from a
disc or a cylinder stretched between the sheets which approach each other.
Branes wrapped
around these cycles are light w.r.t. the Planck mass and decouple from gravity.
Furthermore, it can be shown that they yield
precisely the expected spectrum of a 4D N=2 Yang-Mills theory: massive gauge vector
multiplets
from $S^1 \times S^2$ branes and massive dyon hypermultiplets from $S^3$ branes.
Mathematically, the required degeneration
of a branch of $\Sigma_\alpha$ means that the CY develops a complex curve of singularities.
In our example, there are several  possibilities. The one in which we are interested here
corresponds
to three coinciding sheets, which gives a curve of $A_2$ singularities, and
subsequently $SU(3)$ Yang-Mills. Other possibilities are $SU(2)$ (2
coinciding sheets) and $SU(2) \times SU(2)$ (2 pairs of coinciding
sheets).

Some elementary algebra shows that the desired $SU(3)$ rigid limit is 
obtained by setting
\begin{eqnarray}
B &=& \epsilon  \nonumber\\
A_1 &=& \epsilon^{2/3} u_1 \nonumber\\
A_2 &=& -1 + \epsilon u_2
\end{eqnarray}
and letting $\epsilon \to 0$ while keeping $u_1$ and $u_2$ finite.

The degenerating branch of $\Sigma_\alpha$ is given by the equation
$P(x) = -\sqrt{B^\prime(\zeta)}$.
Rescaling $x = \epsilon^{1/3} \tilde{x}$ yields
\begin{equation}
\tilde{x}^3 + u_1 \tilde{x} + u_2 + \frac{1}{4} (\zeta + \frac{1}{\zeta})
-\frac{1}{2}\epsilon\left( \tilde{x}^3 + u_1 \tilde{x} + u_2 \right)
^2 = 0 \ .\label{SWsurface}
\end{equation}
Putting $\epsilon=0$, this is precisely the genus 2 Seiberg-Witten Riemann surface
for $SU(3)$.

A basis of eight 3-cycles is constructed as follows: two cycles are obtained from
cylinders stretched between the coinciding sheets, two from discs between the coinciding sheets, two
from a cylinder stretched between the sheets of $\Sigma_\beta$ and two from tori encircling
$\Sigma$. The first
four have $\alpha$ as torus cycle, the last four $\beta$.
The leading $\epsilon$ dependence of the corresponding periods can be deduced from the
monodromies of the cycles under $\epsilon \to e^{2 \pi i} \epsilon$. The first four periods
are proportional to $\epsilon^{1/3}$.
The next two periods have a part proportional to $\log \epsilon$ and a part proportional to
$\epsilon^{2/3} \log \epsilon$. The last two have only terms which are analytic in $\epsilon$ or
proportional to $\epsilon^{2/3}$.  Since the mass of a BPS 3-brane is proportional to the
absolute value of the corresponding period, it follows that a brane wrapped around one of the
first
four cycles is much lighter than a brane wrapped around one of the last four.

In fact for the ``light'' periods we can do better and derive, using (\ref{fa}) and (\ref{SWsurface}), an expression to arbitrary order in $\epsilon$:
\begin{equation}
\int_C \Omega = \epsilon^{1/3} \int_{\partial S} \lambda,
\end{equation}
where $S$ is the base surface of $C$, $\partial S$ its boundary, a 1-cycle on the Riemann
surface (\ref{SWsurface}), and $\lambda$ a meromorphic one form given by
\begin{equation}
\lambda = \sqrt{2} \, \frac{\dop \zeta}{\zeta} \, \tilde{x} \,
\left[1 + \frac{\epsilon}{8}\left( \frac{1}{4}\tilde{x}^3 + \frac{1}{2} u_1 \tilde{x} + u_2
- \frac{3}{4}(\zeta + \frac{1}{\zeta})\right) + \cdots\right]\ .
\end{equation}
To lowest order in $\epsilon$, this is exactly the Seiberg-Witten 1-form.

Now the intersection form can be put in block diagonal form
by a (noninteger) redefinition of the last four basis elements. This, together with the a priori not
so obvious fact that the intersection form of the first four 3-cycles is equal to (minus) the intersection
form of the corresponding Seiberg-Witten 1-cycles, guarantees that the argument inside the logarithm
in (\ref{calK}) or (\ref{calK2}) indeed contains a term proportional to the rigid Seiberg-Witten K\"{a}hler potential (\ref{K}). Careful calculation of the other block of the intersection matrix
provides us with the other terms, and we find
\begin{eqnarray}
{\cal K}
&=& -\ln \left( m^2 \ln( k |\epsilon|^{-2})
-  |\epsilon|^{2/3} K_{SW} + O(\epsilon^{4/3})
\right) \nonumber\\
&=& -\ln \left(m^2 \ln(k |\epsilon|^{-2}) \right)
+ \frac{|\epsilon|^{2/3}}{m^2 \ln(k |\epsilon|^{-2})} K_{SW}
+ \cdots\ ,
\end{eqnarray}
where $m^2$ and $k$ are certain constants, $m^2>0$ and $K_{SW}$ is the $SU(3)$ Seiberg-Witten
rigid special K\"{a}hler potential. We conclude that the conditions (\ref{limdef}) are satisfied,
so we indeed have rigid special K\"{a}hler geometry in this limit of 
moduli space. We also
see very clearly now that the field theory degrees of freedom, associated with the dynamics of the four
light branes, are indeed decoupled from gravity.

\paragraph{Acknowledgments.}\par
\noindent
Work supported by
the European Commission TMR programme ERBFMRX-CT96-0045,
in which D.Z. is associated to U. Torino. F.D., W.T. and A.V.P. thank
their employer, the FWO Belgium, for financial support.

\end{document}